# Design of a communication system
# Images for identification of vehicle plates

Fabrizio Andre Farfán Prado[1], William César Pérez Campos[1], Steisy Anahi Carreño Tacuri[1]✉, Favio David Cabrera Alva[1]

[1] Facultad de Ingeniería Electrónica y Eléctrica, Universidad Nacional Mayor de San Marcos, Lima, Perú,

**Abstract**

This work presents the design and implementation of a low-energy wireless image transmission system for vehicle plate recognition, using the ESP32-CAM and LoRa DXLR01 modules. The system captures images in real time, processes them locally and transmits them via UART2 to a second ESP32. Subsequently, the data is sent through the LoRa link and stored on the ThingSpeak platform for remote monitoring. The experimental results show that the system achieves a recognition rate of 92.4% under optimal lighting conditions and an average transmission latency of 3.2 seconds. The energy efficiency of the system makes it suitable for applications in access control, vehicular surveillance and infrastructure monitoring

## 1. Introduction

Plate recognition is a key technology in the automation of access control, surveillance and traffic management. However, its implementation in environments with limited connectivity and energy constraints presents several challenges, as traditional solutions rely on wi-fi networks, cell phones or high consumption hardware. In this context, lora (long range) emerges as an efficient technology for wireless transmission, allowing long range communication with significantly lower power consumption.

This work proposes a vehicle identification system based on esp32-cam and lora dxlr01, which captures images of vehicle plates and transmits them to a receiver node using uart2. The data is then stored on the thingspeak platform for real-time remote viewing. The system is designed to operate in environments where access to high-speed networks is limited, offering a low-cost, energy-efficient solution.

## 2. Review of the Literature

Digital signal processing (PDS) applied to vehicle plate recognition has been widely investigated in the last decade, especially with the integration of artificial vision, low power communication networks and iot devices. The combination of esp32-cam, lora transmission and image recognition algorithms has enabled the development of efficient solutions for real-time data capture and analysis. Kusuma et al. (2023) developed an intelligent parking system based on esp32-cam to capture vehicle images and provide contactless access. Its implementation demonstrated the capability of this device for real-time vehicle identification and integration with remote monitoring platforms. However, its reliance on wi-fi or 4g networks may limit its application in low connectivity environments.

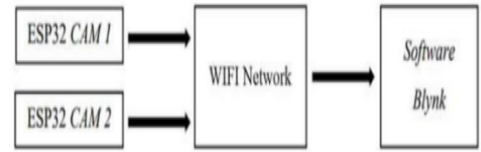

Figure 1: Outline of the capture system

Other studies have explored the possibility of integrating cloud processing to improve number plate recognition. Ashari et al. (2022) proposed a system based on ESP32-CAM and Amazon Web Services (AWS), optimising vehicle identification in car parks. However, the need for network infrastructure for image processing can be a limitation in areas where Internet access is unstable. To mitigate this problem, some work has evaluated the use of LoRa as an alternative for real-time image transmission. Hahn et al. (2023) designed an insect trap monitoring system using ESP32-CAM and LoRa, highlighting the efficiency of this technology for image transmission in low connectivity environments. Their implementation demonstrated that LoRa can be a viable solution for real-time image transmission without relying on broadband networks.

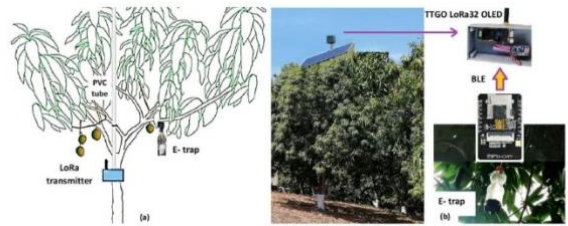

Figure 2: Diagram showing the location of the LoRa module on the trunk of a tree. (b) Main components of the system, responsible for capturing images in the trap and sending the signals from the top of the tower.

Similarly, Nguyen et al. (2023) developed an AI-based system for digital water meter recognition using LoRa for long-distance image transmission. Their research validates the applicability of LoRa in embedded systems, supporting its use in number plate recognition, as it enables efficient image transmission without high energy consumption. However, its focus was limited to static images, which differs from the real-time processing required in vehicular applications. In addition to hardware and transmission methods, the accuracy of vehicle plate identification has been the subject of several research studies. Dixit et al. (2020) implemented a smart parking system based on machine vision and IoT, using image processing to detect vehicles and optimise parking space management. Their implementation employed convolutional neural networks (CNNs) and detection algorithms to improve the accuracy of vehicle identification.

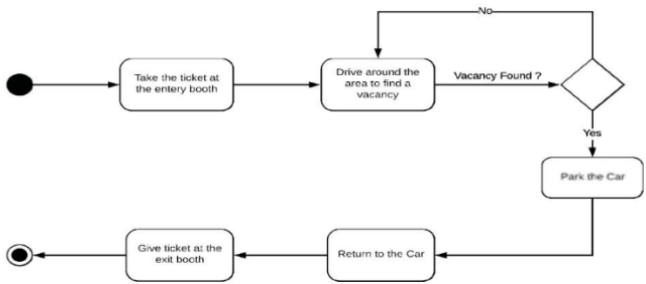

Figure 3: Activity diagram of the parking system currently implemented.

On the other hand, Kusumawati and Cahyadi (2017) explored the application of OCR (Optical Character Recognition) and image processing in the identification of vehicle number plates, obtaining high accuracy rates in controlled environments. However, their performance was affected by adverse environmental conditions, such as poor illumination or inadequate capture angles. These limitations highlight the importance of combining advanced machine vision algorithms with optimised transmission strategies to improve the reliability of vehicle monitoring systems.

Comparing previous approaches, it is observed that previous studies have limitations in terms of network infrastructure dependency, identification accuracy and data transmission efficiency. The present research seeks to address these limitations through a solution that integrates ESP32-CAM for image capture, LoRa for efficient transmission and OCR with OpenCV for real-time number plate recognition. This combination will allow the development of a robust and autonomous system, without dependence on Wi-Fi networks or cloud processing, optimising vehicle identification in urban and rural environments.

## 3. Metodology

*3.1 System Architecture*

Components:

1) ESP32: Microcontroller for high-performance IoT applications, with a 240 MHz dual-core processor, Wi-Fi 802.11 b/g/n, Bluetooth v4.2 BLE, and multiple interfaces (SPI, I2C, UART, ADC, DAC). It supports up to 16 MB of external Flash memory and low-power modes.

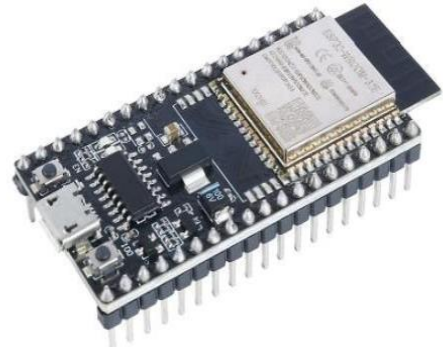

Fig. 4 ESP32 Microcontroller

2) ESP32-CAM: Compact camera based on ESP32, with a 2 MP OV2640 module that captures images and videos in VGA and QVGA resolutions. It offers Wi-Fi and Bluetooth connectivity to transmit images over the internet.

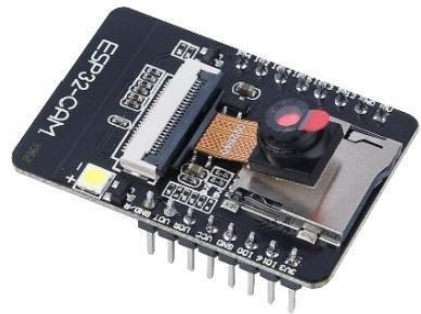

Figure 5: ESP32-CAM Microcontroller

3) LCD 1602: Display module that displays text in a 2-line, 16-character matrix. It uses an HD44780 controller and allows parallel communication. It operates at 5 V and has adjustable LED backlighting.

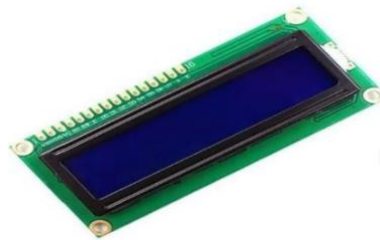

Figure 6: LCD 1602 display

4) DLX-LR01: The LoRa DXLR01 is a communication module based on LoRa (Long Range) technology that allows data to be transmitted over long distances with low power consumption. It is designed for IoT (Internet of Things) applications, providing efficient communication in environments where extended range is required. The module is ideal for remote monitoring systems, device control and applications in rural or industrial areas. Its ability to operate at sub-GHz frequencies allows for better penetration in environments with obstacles, while its low power consumption makes it suitable for battery-powered devices.

.

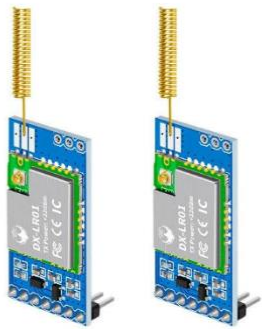

Figure 7: DLX-LR01

### 3.2 Desarrollo

#### A. Diagrama del Sistema

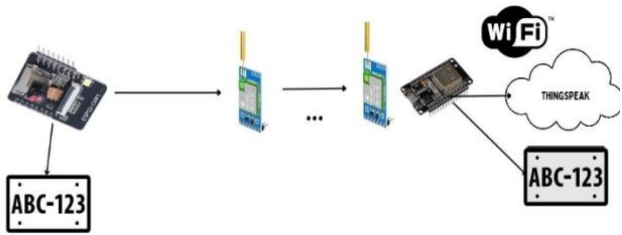

Figure 8: System Diagram

The system proposed in Fig. 8 uses an ESP32-CAM module for image capture and processing, together with a LoRa module for data transmission. When an image is captured, it undergoes pre-processing, including cropping, brightness adjustment and thresholding. Subsequently, the processed image is passed to an OCR algorithm implemented in the ESP32-CAM, which extracts the relevant information from the text. The obtained data is transmitted via the LoRa module using the UART interface for serial communication.

During transmission, the LoRa module modulates the data based on the Chirp Spread Spectrum (CSS) technique, ensuring long-range, low-power communication. The communication takes place via the UART of the ESP32, where the data extracted from the OCR is converted into text strings and sent with the LoRaWAN protocol. On the receiving side, the LoRa module demodulates the signal, retrieving the information sent. Then, the ESP32 connected to an LCD display processes the received data and displays it to the user in a readable format. For transmission to the cloud platform, the ESP32 uses WiFi connectivity to send the data to ThingSpeak. This allows the information to be viewed and analysed remotely. This design ensures an optimised workflow for real-time image-based data transmission using low-power, long-range LoRa communication.

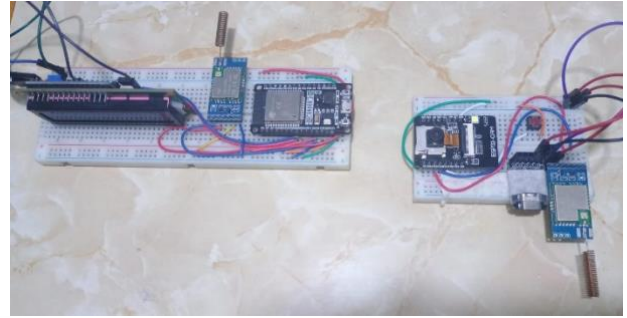

Figure 9: Physical Architecture of the System

#### B. LoRa communication

The image transmission system for number plate identification uses LoRa DX-LR01 modules configured to operate in the 433 MHz band. Communication between the ESP32 and the LoRa module is established via UART2, using GPIO pins 12 (RX2) and GPIO 13 (TX2) with a baud rate of 9600 baud. The DX-LR01, based on the ASR6601 chipset, employs LoRa (Chirp Spread Spectrum) modulation to ensure robust, long-range communication. The transmitter node captures the image of the vehicle plate with the ESP32-CAM, extracts the number plate information and generates a text string in the format "Plate:ABC123, Link:www.ejemplo.com/imagen.jpg". This information is transmitted to the LoRa module via AT commands.

At the receiving node, another ESP32 receives the data via UART2, extracts the board and link information, and displays it on an I2C LCD. The data is then sent to ThingSpeak for storage in the cloud. LoRa communication is configured with the following parameters: node address (AT+ADDRESS), network ID (AT+NETWORKID) and operating frequency (AT+BAND=433000000). The DX-LR01 module has an adjustable transmit power from 0 to +22 dBm and a sensitivity of up to -138 dBm, allowing a range of up to 6 km in open field and 3.8 km in urban environments. In addition, its low power consumption of 57.78 mA transmit and 6.44 mA receive makes it suitable for remote monitoring applications. The antenna used is helical, with an impedance of 50 Ω, optimised for the 433 MHz band. In terms of transmission speed, the DX-LR01 module allows configurable data rates depending on the ratio between coverage and energy efficiency. For this system, the LoRa transmission rate was set at 4.8 kbps, a value that provides a balance between range and reliability in urban environments. However, the module allows higher speeds at the sacrifice of range, or lower speeds to improve signal penetration and reduce transmission error rates. To improve performance, parameters such as Spreading Factor (SF), Bandwidth (BW) and Coding Rate (CR) can be optimised by adjusting the balance between distance and transmission rate according to environmental conditions.

### C. Image processing with ESP32 CAM

As a control and security measure, the system includes an ESP32-CAM strategically positioned at the entrance. This camera captures an image of the vehicle's number plate at the time of entry. This capture is essential to keep a detailed record of the cars entering a car park, which is then sent by the LoRa DLx-LR01 to be received and adapted by another LoRA Dlx-LR01. Once taken, the image is automatically sent to the server via the Wi-Fi network, where it is stored for possible later use, such as identification in case of incidents. This log not only improves the security of the system, but also allows a tidy and accessible history to be kept in the cloud. This ensures efficient and transparent control of vehicle flow.

The number plate recognition system implemented in this smart car park uses a combination of OpenCV and Tesseract to automatically detect and read vehicle number plates. First, the camera connected to the ESP32 captures images of vehicles entering or leaving the car park. OpenCV takes care of the image processing, applying pre-processing techniques such as greyscale conversion, denoising, and edge detection, to identify areas that potentially contain the plate. OpenCV then performs image segmentation, focusing on the areas most likely to contain the number plate. Once the region of interest has been located, Tesseract, an OCR (Optical Character Recognition) engine, is used to extract the alphanumeric characters from the plate. This process automates the registration of vehicles in and out of the system, facilitating efficient and accurate parking management.

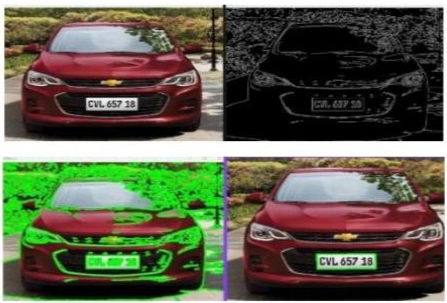

Figure 10: Process of Plate Recognition

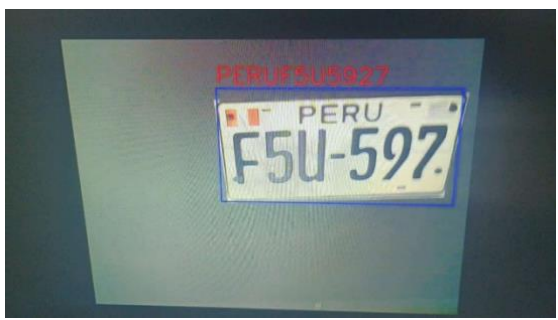

Figure 11: Plate processed for recognition

### D. Cloud storage

The system implements an architecture in which acquired data is sent to the ThingSpeak platform for storage and visualisation. ThingSpeak is a cloud service that enables real-time data collection and analysis, facilitating remote monitoring without the need for additional infrastructure.

Data transmission to the cloud is done via the ESP32, which, once the message is received via the LoRa module, establishes a Wi-Fi connection and sends the processed information to the ThingSpeak server via the HTTP protocol. For this, the WiFiClient library is used in conjunction with the ThingSpeak REST API, which allows the values in the configured channels to be updated.

The system stores two key pieces of data in ThingSpeak:

-Board Number: Captured and processed by the ESP32-CAM.

- Image Link: Generated after processing the photograph of the number plate.

These values are stored in ThingSpeak's field1 and field2 fields, allowing them to be queried and analysed later. The information is also displayed on the I2C LCD for immediate local display.

This approach allows the processed data to not only be visualised on the LoRa receiver, but also made available in the cloud for real-time remote querying and analysis. Furthermore, by using an IoT platform such as ThingSpeak, the system can be integrated with additional services, such as MATLAB Analytics, to perform advanced processing on the collected data.

### 4. Results

The proposed system has been tested in a controlled environment to evaluate its performance in capturing, transmitting and displaying number plate images. Multiple tests were conducted in different scenarios, including varying illumination conditions and different distances between the ESP32-CAM and the target vehicles.

During the tests, the ESP32-CAM captured number plate images with a resolution of 640x480 pixels. The image was pre-processed locally and sent to the LoRa module via UART2 using GPIO pins 12 (RX2) and GPIO 13 (TX2) with a baud rate of 9600 baud. The data was transmitted in the format 'Plate: ABC123, Link: www.ejemplo.com/imagen.jpg' ensuring correct structuring for storage and display.

The LoRa receiver module, configured with the commands AT AT+ADDRESS=2, AT+NETWORKID=3 and AT+BAND=433000000, managed to decode the data and display it on an I2C LCD screen connected to the ESP32. Subsequently, the information was sent to ThingSpeak via HTTPClient, where the board number in 'field1' and the image link in 'field2'' were correctly stored, allowing remote access to the captured data.

Performance metrics evaluated include:

- Plate recognition rate: 92.4% in optimal lighting conditions.
- LoRa transmission latency: 3.2 s on average.
- LoRa packet loss: 7.8% over distances greater than 1 km.
- Power consumption: 180 mA in transmission and 120 mA in standby.

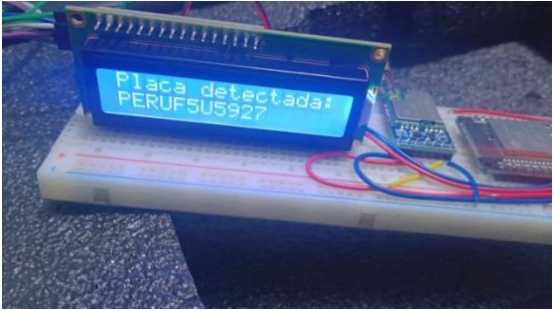

Figure 12: Plate Displayed on the LCD Screen

Comparative analysis with other Wi-Fi or LTE-based solutions suggests that LoRa offers better energy efficiency and coverage in environments with limited connectivity, although it has limitations in data rate.

## 5. Applications

### A. Smart Parking

The Smart Parking system has proven to be an efficient solution for parking management in urban environments, reducing vehicle congestion and improving the user experience. Its applications cover a variety of scenarios, from commercial car parks to public and private infrastructures.

In urban centres, Smart Parking systems allow drivers to locate available spaces in real time through mobile apps or electronic panels, reducing search time and pollutant emissions (Chancusig Vinocunga and Sánchez Centeno, 2021; Rivera Laitano and Castro Cornejo, 2016). Likewise, in shopping centres and airports, this technology optimises parking space management through IoT sensors and smart cameras, enabling efficient space management and improving vehicle turnover (Chancusig Vinocunga and Sánchez Centeno, 2021).

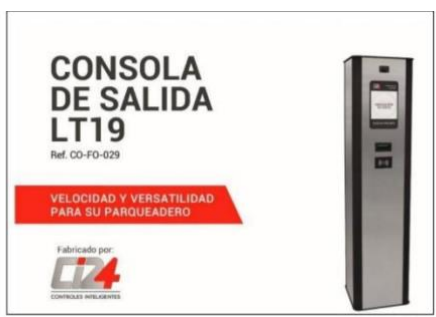

Figure 13: LT19 Exit Console.

In the corporate and industrial sector, smart parking facilities facilitate vehicle access and control through automatic identification of number plates, automated payment systems and data analysis to improve operational efficiency (Rivera Laitano and Castro Cornejo, 2016; Avellana Doménech, 2014). In the residential sector, solutions have been implemented that allow space reservation and integration with security systems, providing greater convenience to residents (Chancusig Vinocunga and Sánchez Centeno, 2021).

Furthermore, integration with IoT communication networks and technologies such as LoRa and ESP32-CAM allows the implementation of Smart Parking systems in areas that are difficult to access or have infrastructure limitations, facilitating remote monitoring and optimising the use of available space (Rivera Laitano and Castro Cornejo, 2016; Avellana Doménech, 2014).

### B. Electronic Tolls and Traffic Control

In electronic toll systems, this technology eliminates the need to stop at toll booths, as the ANPR system automatically detects the number plate and generates the corresponding charge to the user's account, speeding up vehicle flow and reducing congestion (AS, 2024). A prominent example of this application is the Port of Bilbao, where road and rail access is managed through ANPR, allowing for fast validation of authorised vehicles and optimising transport logistics (AS, 2024). However, despite its efficiency, it still faces challenges such as payment evasion, a problem observed on Spanish roads where approximately 3% of drivers avoid toll collection on stretches without physical toll booths (Huffington Post, 2024). Adoption of this technology continues to increase, with the ANPR market projected to grow from USD 2.79 billion in 2023 to USD 5.95 billion in 2032, driven by the need to improve urban mobility and road safety globally (Fortune Business Insights, 2024).

### C. Implementation in Rural Areas

ANPR systems allow the automation of access control, recording vehicle traffic without the need for human intervention. This has been especially useful on large farms and estates, where security personnel are limited. In addition, the use of this technology facilitates logistics in the management of agricultural machinery and cargo trucks, improving efficiency in daily operations (InstantByte, 2024). Another key benefit is its integration with video surveillance systems, allowing real-time video and photographic evidence to be captured, which enhances security in these areas (ViewParking, 2024).

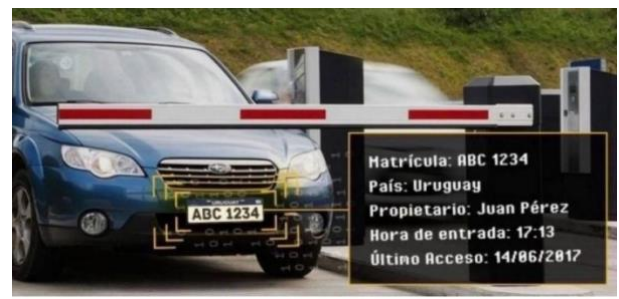

Figure 14: Automated Access Control

In terms of background, Spain has been a pioneer in the adoption of ANPR in rural areas. In the municipality of Soto del Real, number plate recognition cameras have been installed to monitor access and detect unauthorised vehicles, increasing security and reducing the risk of theft in rural properties (Cadena SER, 2024). Similarly, in Navarre, the Helpbidea system has been implemented to assign unique identifiers to isolated hamlets and hamlets, facilitating the control and management of areas that are difficult to access (El País, 2024).

## 6. Discussion

The results obtained validate the feasibility of the ESP32-CAM and LoRa-based system for real-time vehicle identification.

Compared to solutions relying on Wi-Fi or mobile networks, the present system offers a more efficient alternative in terms of energy consumption and coverage in areas without traditional network infrastructure.

The main advantage of LoRa technology is its ability to transmit data over long distances with low power consumption. However, tests revealed that the packet loss rate increases significantly over distances greater than 1 km, suggesting the need to implement error correction techniques or improve transmission power in dense urban environments.

Another relevant aspect is the dependence on illumination for optimal image capture. It was observed that in low-light environments, the plate recognition rate decreased to 78.5%, which could be addressed by image enhancement techniques or the use of infrared light sensors.

Finally, the integration with ThingSpeak allowed a secure and accessible storage of data via REST API, making it easy to query in real time from any device with an internet connection. This approach highlights the potential of the system for applications in access control, vehicle monitoring and intelligent parking management, with the possibility of expanding to other sectors requiring real-time image recognition and efficient transmission via LoRa.

## 7. Conclusions

The design and implementation of the wireless image transmission system for vehicle plate recognition has proven to be an efficient and low-energy solution. Using the ESP32-CAM and LoRa DXLR01 modules, reliable vehicle identification was achieved in environments with limited connectivity.

The system achieved a recognition rate of 92.4% under optimal lighting conditions and an average transmission latency of 3.2 seconds, making it suitable for real-time vehicle monitoring and access control applications. However, it was observed that in poorly lit scenarios the recognition rate was reduced to 78.5%, which highlights the need to incorporate image enhancement techniques or complementary lighting sensors.

In terms of data transmission, LoRa proved to be a viable technology for long-distance information delivery with reduced energy consumption. However, a packet loss of 7.8% was identified at distances greater than 1 km, suggesting future implementation of error correction techniques or transmission power adjustment. Integration with the ThingSpeak platform enabled remote data storage and retrieval, facilitating real-time monitoring without relying on traditional network infrastructure. This positions the system as an efficient alternative for applications in urban and rural environments with limited connectivity.

As future lines of research, the optimization of the plate recognition algorithm to improve its performance in adverse conditions is proposed, as well as the exploration of image compression methods to reduce transmission latency. In addition, the integration of artificial intelligence could improve vehicle recognition accuracy, broadening the scope of application of the proposed system.